\documentstyle[pra,aps,epsfig]{revtex}
\begin{document}
\draft
\title{The Quantum Double Pendulum:\\
a Study of an Autonomous Classically Chaotic Quantum System}
\author{Luca Perotti}
\address{Center for Nonlinear and Complex Systems, Universit\'a degli studi dell'Insubria, Como}
\date{\today}
\maketitle

\begin{abstract}
A numerical study of the quantum double pendulum is conducted. A suitable quantum scaling is found which allows to have as the only parameters the ratios of the lengths and masses of the two pendula and a (quantum) gravity parameter containing Planck's constant. Comparison with classical and semiclassical results is used to understand the behaviour of the energy curves of the levels, to define regimes in terms of the gravity parameter, and to classify the (resonant) interactions among levels by connecting them to various classical phase space structures (resonance islands). 
\end{abstract}


\section{Introduction}

In 1992, J. Ford suggested the double pendulum as a suitable system on which to experimentally test the ability of quantum mechanics to describe classically chaotic systems \cite{ford}: as a simple spatially bounded autonomous Hamiltonian system presenting -in its classical version- transition to chaos it appeared ideal to test the consequences of the lack of chaos in ``eigenfunctions, eigenvalues and time evolution" of such quantum systems. Since then, our understanding of the signatures of classical chaos in quantum mechanics has greatly improved \cite{chao}; in particular arguments have been advanced to the effect that continuous observation of the quantum system would bring its behaviour back to the classical one \cite{cont}. Still the above mentioned characteristics of the double pendulum  make it a candidate for a thorough quantum study: it is far simpler than other chaotic autonomous systems that have been studied in recent years, like quantum stadia \cite{stad}, whith their relevance to quantum dot technology, or hydrogen in strong magnetic fields \cite{diama} and helium \cite{elio}, which moreover are not spatially bounded; the quantum kicked rotor \cite{rota} and hydrogen in monochromatic microwave fields \cite{idro} have been the focus of much study and have almost become a paradigm of ``quantum chaos", but they are non-autonomous.

The relative freedom with which mass and lenght ratios can be changed in the double pendulum would moreover allow us to study how the classical transition to chaos is reflected in its quantum counterpart when varying these parameters and this may be illustrative of various ``quenchings" of chaos in systems like for example helium itself: the absence of observed chaos in helium appears to be a consequence of the possibility of an adiabatic seperation in hyperspherical coordinates \cite{elio} so that there are quite good quantum numbers right up to the present precision of observations. Different mass ratios would probably break this quasi-symmetry; but changing the electron masses would just result in making the system even more complicated and not accessible to direct experimental testing.  

On the other hand, even though the classical double pendulum has often been used as an example of autonomous chaotic system \cite{cla,vari}, doubts about the proper ordering of the non-commuting operators in its kinetic energy \cite{ali} have up to now made its quantum counterpart not palatable to extensive analysis. The aim of my paper is to propose a reasonable ordering giving the proper behaviours for zero gravity (the behaviour in the high gravity limit does not depend on the ordering), and to explore the properties of the eigenfunctions of the resulting Hamiltonian. To my knowledge this is the first time that the quantum dynamics of the double pendulum is analyzed at all values of gravity and not only in the high gravity limit where it reduces to the trivial case of two coupled harmonic oscillators.

The present paper is thus organized: section \ref{uno1} presents the system, both classical and quantum; section \ref{due} introduces and discusses the quantum numerical methods used. Finally, section \ref{tre} discusses the quantum system properties; first those that can be obtained by semiclassical methods, then those obtained from the quantum simulations, focusing on the different quantum behaviour in the three classical regimes of low gravity (classical regular motion in most of the phase space), medium gravity (classical ``global chaos" regime) and high gravity (regular ``coupled oscillators" regime).

\section{The Model: Classical and Quantum Hamiltonians}\label{uno1}

\begin{figure}[htbp]
\centering\epsfig{file=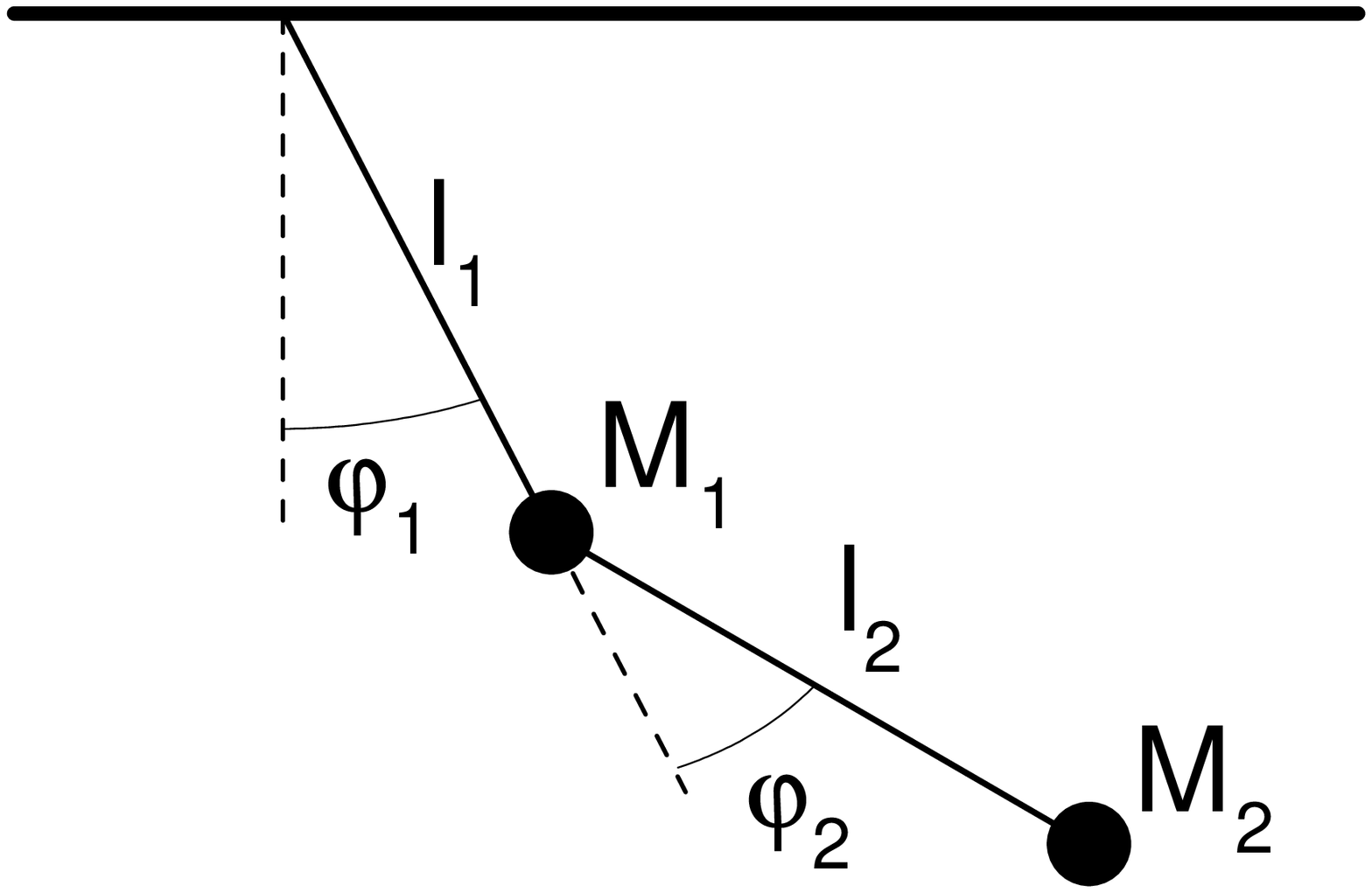,width=1.0\linewidth}
\caption{An ideal double pendulum.}
\label{fig1}
\end{figure}

An ideal double pendulum is shown in fig. \ref{fig1}; $l_1$ and $l_2$ are the lengths of the two pendula and $M_1$ and $M_2$ their masses. Introducing the ratios $l=l_2/l_1$ and $\mu=M_2/M_1$ and the momenta $L_1$ (total angular momentum of the system) and $L_2$ conjugated to the two angles $\varphi_1$ and $\varphi_2$, the classical Hamiltonian reads \cite{cla}:
\begin{eqnarray}
H={1\over{2M_1 l_1^2}}\left\{{{L_1^2}\over{1+\mu \sin^2{\varphi_2}}}-
{{2L_1L_2}\over l}\left[{{l+\cos{\varphi_2}}\over{1+\mu \sin^2{\varphi_2}} }\right]+ 
 L_2^2{{1+\mu+2\mu l\cos{\varphi_2}+\mu l^2}\over{\mu l^2(1+\mu \sin^2{\varphi_2})}} \right\}+\nonumber\\
M_1 g l_1 \left\{(1+\mu)(1-\cos{\varphi_1})+\mu l [1-\cos{(\varphi_1+\varphi_2)}]\right\} 
\end{eqnarray}

It is possible \cite{cla} to scale the system via the adoption of new adimensional variables: time $\tau= t\sqrt{E/(M_1 l_1^2)}$ and momenta $\lambda_i = L_i/\sqrt{E M_1 l_1^2}$, where $E$ is the constant total energy. The scaled Hamiltonian $h=H/E$ then always equals $1$ and depends only on the gravity parameter $\gamma = g M_1 l_1/E$ and the ratios $l$ and $\mu$: 
\begin{eqnarray}
h= {1\over 2}\left\{{{\lambda_1^2}\over{1+\mu \sin^2{\varphi_2}}}-
{{2\lambda_1\lambda_2}\over l}\left[{{l+\cos{\varphi_2}}\over{1+\mu \sin^2{\varphi_2}} }\right]+ 
 \lambda_2^2{{1+\mu+2\mu l\cos{\varphi_2}+\mu l^2}\over{\mu l^2(1+\mu \sin^2{\varphi_2})}} \right\}+\nonumber\\
\gamma \left\{(1+\mu)(1-\cos{\varphi_1})+\mu l [1-\cos{(\varphi_1+\varphi_2)}]\right\}.\label{rid} 
\end{eqnarray}

Classical results are presented through Poincar\'e sufaces of section (SOS); in the present paper I shall only consider sufaces of section in the $\varphi_2=0$, $\dot{\varphi}_2>0$, $\{\lambda_1,\varphi_1\}$ plane of which several examples are given in Fig. $8$ of Ref. \cite{cla}. 

For $\gamma=0$ the total angular momentum $\lambda_1$ is conserved, it is therefore one of the two actions of the system and the SOS consists of horizontal lines; motion in $\varphi_1$ is always a rotation, but -as $\varphi_1$ is not the angle associated to that action- $\dot{\varphi_1}$ is not constant. The second action $I_2$ has instead to be calculated numerically \cite{cla}. The uniformity of the SOS also hides the two different kinds of motion in $\varphi_2=0$: rotation for $|\lambda_1|<\sqrt{2[1+\mu(1-l)^2]}$, and libration for $\sqrt{2[1+\mu(1-l)^2]}<|\lambda_1|<\sqrt{2[1+\mu(1+l)^2]}$.

For $ \gamma \neq 0$ we shall here only note two facts. One is the vertical asymmetry of the SOS which is due to the fact that -following Poincar\'e prescription- only the orbits crossing it with $\dot{\varphi_2} >0$ are shown; the SOS for $\dot{\varphi_2} <0$ is perfectly symmetric to it. The other is that for $\gamma> [2(1+\mu+\mu l)]^{-1}$ rotation in $\varphi_1$ is no longer possible and $\varphi_1$ is limited between $\pm \arccos{(1- [\gamma(1+\mu+\mu l)]^{-1})}$.

In quantum mechanics we cannot use the classical scaling; we instead multiply the Hamiltonian by $2 M_1 l_1^2/\hbar^2$ so as to have as sole parameters the scaled (adimensional) gravity $\tilde{\gamma}=2M_1^2 l_1^3 g/\hbar^2$ and again the two ratios $l$ and $\mu$; the adimensional scaled energy will be indicated as $\tilde{E}=2M_1 l_1^2 E/\hbar^2$, the time as $\tilde{t}=t\hbar/2M_1 l_1^2$, and the adimensional scaled momentum operators as $\hat{\tilde{L}}_i = \hat{L}_i/ \hbar$ where the quantum momentum operators are defined in the usual way:
$\hat{L}_i=-i\hbar {{\partial}/{\partial \varphi_i}}$. The symmetrized quantum Hamiltonian then reads
\begin{eqnarray}
\tilde{H}= \left\{{{\hat{\tilde{L}}_1^2}\over{1+\mu \sin^2{\varphi_2}}}-
{{\hat{\tilde{L}}_1}\over l}\left[{{l+\cos{\varphi_2}}\over{1+\mu \sin^2{\varphi_2}}}\hat{\tilde{L}}_2+
\hat{\tilde{L}}_2{{l+\cos{\varphi_2}}\over{1+\mu \sin^2{\varphi_2}}}\right]+\right.\nonumber\\
\left. \hat{\tilde{L}}_2{{1+\mu+2\mu l\cos{\varphi_2}+\mu l^2}\over{\mu l^2(1+\mu \sin^2{\varphi_2})}}\hat{\tilde{L}}_2\right\}+
\tilde{\gamma} \left\{(1+\mu)(1-\cos{\varphi_1})+\mu l [1-\cos{(\varphi_1+\varphi_2)}]\right\}\label{sch}
\end{eqnarray}

Infinitely many other symmetrizations of the last two kinetic terms are possible \cite{note}; our choice has been dictated by the physical argument that for zero gravity the ground state is completely delocalized \cite{dubbio}; its energy must therefore be zero. As we shall see, the chosen symmetrization guarantees that this be the case, even for the truncated basis sets we have to use for our numerical simulations.  

Classical and quantum adimensional scaled parameters and variables are related thus: 
\begin{eqnarray}
\gamma = \tilde{\gamma}/\tilde{E};\label{grav}\\ 
\lambda_i = \tilde{L}_i \sqrt{2/\tilde{E}}
\label{act}\\
\tau = \tilde{t} \sqrt{2\tilde{E}}.
\end{eqnarray}
Eq. (\ref{grav}) means that -for given values of $l$ and $\mu$- the energy levels corresponding to the same classical situation, as described by the SOS at a given classical gravity parameter $\bar{\gamma}$, are to be found on the $\{\tilde{\gamma},\tilde{E}\}$ plane along the straight line $\tilde{E} = \tilde{\gamma} / \bar{\gamma}$. 

The classical limit is obtained for $\tilde{\gamma} \to \infty$ along such a line; this means having $M_1,l_1 \to \infty$ while keeping all three the classical parameters $l, \mu$, and $\bar{\gamma}$ constant. The same result is obtained with the usual ``unphysical" limit $\hbar \to 0$.

\section{Numerical Methods}\label{due}

\subsection{The Projection of the Double Pendulum on a Rotor Basis}

To numerically calculate the energy levels of the Hamiltonian (\ref{sch}) we project it on the basis $\Phi_{ m_1 m_2 }$ given by the tensor product of the bases for two free rotors:
\begin{eqnarray}
\Phi_{ m_1 m_2 }={{e^{i m_1 \varphi_1 } e^{i m_2 \varphi_2} }\over{2 \pi}};
\end{eqnarray}
and then we diagonalize the finite matrix obtained by truncating the basis at suitables values of the indices $m_1$ and $m_2$ \cite{note4}. 
The matrix elements $\left< m'_1 m'_2 | \tilde{H} |m_1 m_2 \right>= \int_0^{2\pi}\int_0^{2\pi}{d \varphi_1 d \varphi_2 \Phi^*_{ m'_1 m'_2 }\tilde{H} \Phi_{ m_1 m_2 }}$ are:
\begin{eqnarray}
\left< m'_1 m'_2 | \tilde{K} |m_1 m_2 \right>= 
\left\{{1\over{\sqrt{1+\mu}}}\left[m_1(m_1-m_2-m'_2)+m_2m'_2{{1+\mu+\mu l^2}\over{\mu l^2}}\right]\left({{2+\mu-2\sqrt{1+\mu}}\over{\mu}}\right)^{|n|}\delta_{m_2 ,m'_2+2n}+\right.\label{uno}\\
\left. +{{\sqrt{1+\mu}-1}\over{\mu l}}\left[ 2m_2m'_2-m_1(m_2+m'_2) \right]\left({{2+\mu-2\sqrt{1+\mu}}\over{\mu}}\right)^{g(n)}\delta_{m_2 ,m'_2+(2n+1)} \right\}\delta_{m_1 ,m'_1}\nonumber\\
g(n)=  \hspace{0.1in} |n| \hspace{1.1in} n\geq 0 \nonumber \\
g(n)=  |n|-1 \hspace{1.0in} n < 0\nonumber\\
\left< m'_1 m'_2 | \tilde{U }|m_1 m_2 \right>=\tilde{\gamma}\left\{(1+\mu)\delta_{m_2 ,m'_2}\left[\delta_{m_1 ,m'_1}-{1\over 2}(\delta_{m_1 ,m'_1-1}+\delta_{m_1 ,m'_1+1})\right]+\right.\hspace{1.9in}\\
\left. +\mu l
\left[\delta_{m_2 ,m'_2}\delta_{m_1 ,m'_1}-{1\over 2}(\delta_{m_2 ,m'_2-1}\delta_{m_1 ,m'_1-1}+\delta_{m_2 ,m'_2+1}\delta_{m_1 ,m'_1+1})\right]\right\}\hspace{1.0in}\nonumber
\end{eqnarray}
where we have separated the kinetic energy term $\tilde{K}$ and the potential energy one  $\tilde{U }$.

It can immediately be seen that the kinetic energy matrix elements between the state $m_1=m_2=0$ and every basis state (including itself) are zero, thus giving zero as an eigenvalue of the system for zero gravity. 

When running a simulation on a trucated basis, it is important to evaluate how many states are a good approximation to those of the full problem. Our simulations for $\mu=l=1$ use $|m_1|^{max}=25$ ($51$ levels) and $|m_2|^{max}=18$ ($37$ levels) \cite{note3}; a test run at $\tilde{\gamma}=0$ with a doubled basis set ($81\times 49$ is the best choice in this case) shows that $555$ levels ($\tilde{E}^{max} = 145$) are practically identical to those calculated on the smaller basis, giving a total of about $30\%$ of reliable states. With increasing $\tilde{\gamma}$ this number decreases, since the lack of interaction with the missing states at high energies will progressively make also the topmost reliable states unreliable. On the other hand, since the energy of all levels grows with $\tilde{\gamma}$, $\tilde{E}^{max}$ grows at least as the slowest growing state, namely the ground state which grows as $\sqrt{\tilde{\gamma}}$. 

\begin{figure}[htbp]
\centering\epsfig{file=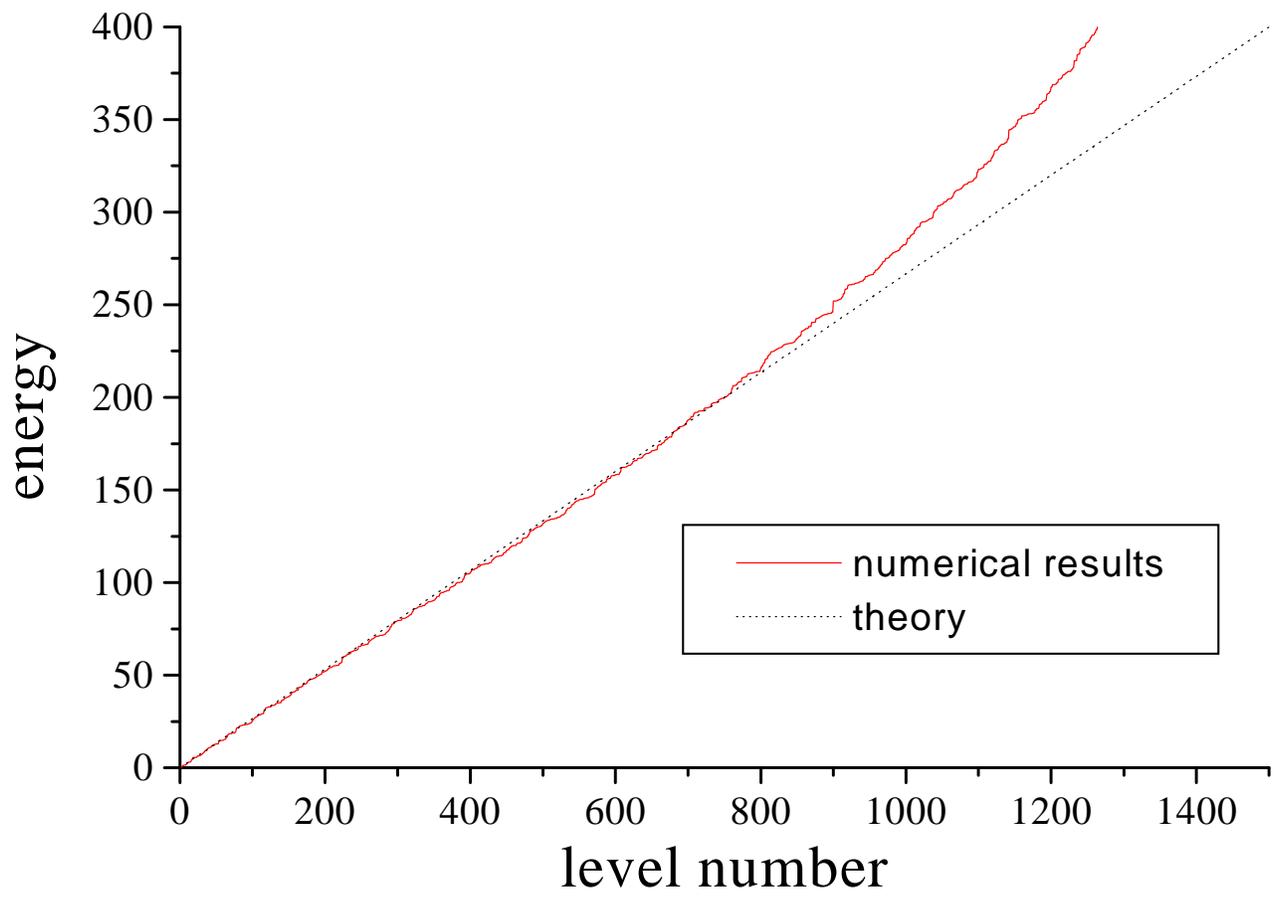,width=1.0\linewidth}
\caption{Comparison of theoretical (dash) and numerical (full line) energy Vs. level number curves.}
\label{fig3}
\end{figure}

We have performed other two reliability tests on (ordering independent) properties of the system:

1) We confronted the $l=\mu=1, \gamma=0$ level density with the theoretical value obtained from the third graph in Fig. \ref{fig2} following the procedure given in section \ref{class}. As shown in Fig. \ref{fig3}, the agreement is very good for all the states we have found above to be reliable.

2) Again for $\mu=l=1$, the dependence of the ground state energy from $\tilde{\gamma}$ approaches for high values of $\tilde{\gamma}$ the theoretical one from eq. (\ref{ene}): $\tilde{E}_{0,0}= \sqrt{\tilde{\gamma}(2+\sqrt{2})}\approx 1.848\sqrt{\tilde{\gamma}}$.

\subsection{Husimi Functions}

For many years now, Husimi functions \cite{tab} have been widely used when comparing quantum and classical systems, as they allow to project quantum functions in phase space in a way that avoids the interpretation problems connected with Wigner functions.
In Hilbert space, the coherent states to be used as coarse-graining functions are, for the cylindrical phase space of each of the two spatial variables $i=1,2$, and apart from unnecessary constant phase terms \cite{coe},
\begin{eqnarray}
\left|\Psi_{\bar{\varphi}_i,\tilde{\bar{L}}_i}\right>=\sqrt{1\over{2\sigma_i\pi^{3/2}}}
\Sigma_{m_i} e^{-{1\over{2\sigma_i^2}}(m_i-\tilde{\bar{L}}_i)^2-i\bar{\varphi}_i m_i}\left|m_i\right>.\nonumber
\end{eqnarray}
Here the variables used are the adimensional quantum scaled ones, $\bar{\varphi}_i$ and $\tilde{\bar{L}}_i$ are the (quantum scaled) phase space coordinates of the center of the packet, and $\sigma_i$ -the angular momentum width parameter- is a free parameter; good results are obtained when $\sigma\simeq 1$.
The normalization is chosen so that the Husimi function of any single rotor eigenstate $\rho_H^{(m_i)}=\left|\left<m_i\right|\left.\Psi_{\bar{\varphi}_i,\tilde{\bar{L}}_i}\right>\right|^2$ is normalized to $1$. 

The Husimi function for a double pendulum eigenstate $\left|\Phi\right>= \Sigma_{m_1,m_2} C_{m_1,m_2}\left|m_1,m_2\right>$ will therefore be:
\begin{eqnarray}
\rho_H=\left|\left<\Psi_{\bar{\varphi}_1,\tilde{\bar{L}}_1,\bar{\varphi}_2,\tilde{\bar{L}}_2}\right|\left.\Phi\right>\right|^2={1\over{4\sigma_1\sigma_2\pi^3}}
\left|\Sigma_{m_1,m_2} C_{m_1,m_2} e^{-{1\over{2\sigma_1^2}}(m_1-\tilde{\bar{L}}_1)^2-{1\over{2\sigma_2^2}}(m_2-\tilde{\bar{L}}_2)^2+i(\bar{\varphi}_1 m_1+\bar{\varphi}_2 m_2)}\right|^2\label{hu}
\end{eqnarray}
For comparison with the classical Poincar\'e surfaces of section, we shall here calculate only the Husimi function (\ref{hu}) on the surface $(\lambda_1, \varphi_1)$, where $\varphi_2=0$ and 
\begin{eqnarray}
\tilde{\bar{L}}_2={{(1+{1\over l})\tilde{\bar{L}}_1+\sqrt{[\tilde{E}-\tilde{\gamma}(1+\mu+\mu l)(1-\cos \bar{\varphi}_1)]\left({{1+\mu+2\mu l+\mu l^2}\over{\mu l^2}}\right)-{{\tilde{\bar{L}}_1^2}\over{\mu l^2}}}}\over{{1+\mu+2\mu l+\mu l^2}\over{\mu l^2}}},\nonumber
\end{eqnarray}
so that $\dot{\varphi}_2>0$.
 
\section{Results}\label{tre}

Now that we have the necessary numerical tools, we can use them to explore the quantum behaviour in the three classical regimes which we encounter when increasing the classical gravity parameter $\gamma$ from $0$ to $\infty$ \cite{cla}: regular motion in most of the phase space ($\gamma \sim 0$), ``global chaos" regime, and regular ``coupled oscillators" regime ($\gamma \gg \max(1/(2\mu l) , 1/[2(1+\mu)])$). First though I shall make some general considerations which will help orientate us in the parameter space, and then pass to a detailed analysis of my numerical results.

\subsection{Level classification and densities}\label{class}

Energy levels at a given $\tilde{\gamma}$ can be classified in three groups according to the character of the classical SOS they correspond to. Starting from the bottom we first have ``coupled harmonic oscillators states" (high $\gamma$), then ``chaotic" states (medium $\gamma$), and finally ``free rotors" states (low $\gamma$).

For low $\tilde{\gamma}$'s only few states belong to the first two classes; for increasing $\tilde{\gamma}$, their number grows, but it remains finite for any finite value of $\tilde{\gamma}$; the number of free rotors states is instead infinite for every value of $\tilde{\gamma}$. 

For ``coupled harmonic oscillators" states (in whose number is included the ground state), the energy levels $\tilde{E}_{n_1,n_2}$ are given by the expression
\begin{eqnarray}
\tilde{E}_{n_1,n_2}= \sqrt{2\tilde{\gamma}}\left[\alpha_1\left(n_1+{1\over 2}\right)+\alpha_2\left(n_2+{1\over 2}\right)\right]\label{ene}
\end{eqnarray}
and therefore grow as the square root of $\tilde{\gamma}$. In (\ref{ene}) the frequency factors are \cite{cla}
\begin{eqnarray}
\alpha_{1,2}= \sqrt{{(1+\mu)(1+l) \pm \sqrt{(1+\mu)^2(1+l)^2-4l(1+\mu)}}\over {2l}},\nonumber
\end{eqnarray}
and the Maslov indices are both $1/2$, as each of the two oscillators has two caustics (in this case the inversion points on the paths on which the actions are calculated) \cite{tab}.

The number of levels under a given value $\tilde{\bar{E}}$ of the energy is therefore
\begin{eqnarray}
N\simeq  {{\tilde{\bar{E}}^2/2}\over {2\tilde{\gamma}\alpha_1 \alpha_2}}={{\tilde{\bar{E}}^2}\over {2\tilde{\gamma}}}\sqrt{{1+\mu}\over l}, \label{numer}
\end{eqnarray}
so that, on one hand, the density of levels is:
\begin{eqnarray}
{{dN}\over{d\tilde{E}}}\simeq {{\tilde{E}}\over {\tilde{\gamma}}}\sqrt{{1+\mu}\over l}={1\over {\gamma}}\sqrt{{1+\mu}\over l}.\label{den}
\end{eqnarray}
and, on the other, the number of levels under a given value $\bar{\gamma}$ of the classical gravity parameter $\gamma$  grows linearly with $\tilde{\gamma}$: 
\begin{eqnarray}
\bar{N}\simeq {{\tilde{\gamma}}\over {4\bar{\gamma}^2\alpha_1 \alpha_2}}={{\tilde{\gamma}}\over {2\bar{\gamma}^2}}\sqrt{{1+\mu}\over l} .\nonumber
\end{eqnarray}

For ``free rotors" states ($\tilde{\gamma}=0$), it is more difficult to exactly evaluate the density of levels, as the second action cannot be calculated analytically. On the other hand some considerations can be made: while in ``coupled harmonic oscillators" regime the energy is directly proportional to the scaled actions $n_i$ (see eq. \ref{ene}), in ``free rotors" regime the energy is proportional to the actions squared (see eq. \ref{act}). Since the number of levels below a given energy is proportional to the product of the actions, in the latter case it is only linear in energy (as opposed to the quadratic dependence we have in the former case: see eq. \ref{numer}) and the density of states is constant. This constant has to be evaluated numerically: to do it we start by plotting the classical scaled action $I_2$ versus $I_1=\lambda_1$ thus obtaining the constant energy curve at the classical scaled energy $h$ (that by definition equals $1$). We now note that increasing the energy the graph expands radially. the total number of states below a given energy $\tilde{\bar{E}}$ is therefore the number of couples of quantized actions that can fit in the area ${\mathcal A}_{\tilde{\bar{E}}}$ swept by the constant energy curve in its growth from $\tilde{E}=0$ to $\tilde{E}=\tilde{\bar{E}}$ or -equivalently- the area swept by a radius connecting the points of the $\tilde{E}=\tilde{\bar{E}}$ curve to the origin. Care must be taken to calculate twice the areas swept both by the outer curves (corresponding to libration in the second angle $\varphi_2$) and by the inner ones (rotation in $\varphi_2$), as regions of phase space corresponding to different classes of motion have different quantum numbers \cite{note5}. We now recall the relationship eq. \ref{act} between classical and quantum scaled actions that here reads $n_i= I_i \sqrt{{\tilde{E}}/ 2}$, where $n_1=m_1$ and $n_2$ are the quantum numbers at energy $\tilde{E}$ and $I_i,\hspace{.05in} i=1,2$ are again the classical scaled actions; the level density $dN / d\tilde{E}$ therefore equals half of the area $2{\mathcal A}_h$ as measured for the classical scaled energy $h=1$. Examples are given in Fig. \ref{fig2}.  

In quantum scaled variables, the energy of the lowest states grows as $\sqrt{\tilde{\gamma}}$; the energy of the highest states instead grows linearly with $\tilde{\gamma}$; the net result is a decrease in the density of states with $\tilde{\gamma}$ at any given energy $\tilde{E}$.
On the other hand, when going to the classical limit ($\tilde{\gamma}\to \infty$, $\gamma=const$), the density of states remains constant both because of the constant density at $\tilde{\gamma}=0$ (for low $\gamma$) and because of eq. \ref{den} (for high $\gamma$).  

In terms of physical variables the energy density instead grows going to the classical limit, due to the relationship ${{dN}\over{dE}}= {{2M_1l_1^2}\over {\hbar^2}}{{dN}\over{d\tilde{E}}}$, but this contribution is a uniform scale one: it does not alter the level structure.

\subsection{Level Interaction and Relationship between Husimi Functions and Classical SOS for $l=\mu=1$}

\subsubsection{Level Structure and General Considerations}

Fig. \ref{fig4} shows the energy curves for $\tilde{E}$ up to $60$ and $\tilde{\gamma}$ up to $10$.
As expected from our discussion in the previous section, we see that the lowest levels -which almost from the start are in the ``coupled harmonic oscillators" regime- grow as $\sqrt{\tilde{\gamma}}$; most of the other levels instead grow at first linearly with $\tilde{\gamma}$. This is a consequence of the adiabatic teorem for noninteracting levels: as the action is approximately constant, the growth in energy of these levels goes as the average potential energy $<U>=\tilde{\gamma}(1+\mu+\mu l)= 3\tilde{\gamma}$. Thus the levels exibiting such behaviour must be (as confirmed by their Husimi functions) those associated at first with the surviving KAM tori \cite{lib} and then with the island chains with long recurrence times which are located at the highest values of $|\lambda_1|$ and at $|\lambda_1|<\sim 0$; in particular we shall see that the levels associated with the KAM tori at the highest values of $|\lambda_1|$ are very resistent and preserve the shape of their Husimi functions well into the ``global chaos" region, where -if we look at any classical SOS- we see chaos almost everywhere.
Only when the growth in energy of these levels slows down do the Husimi functions change.

On the other hand, the levels associated with resonance islands grow from the start more slowly than the others \cite{idro1}. Extremely noticeable are the groups of levels associated with the main resonance island just below the positive branch of the separatrix ($\lambda_1 = \sqrt{2}$); those starting at $\tilde{E}= 1.5, 5.5, 12.0, 21.5, 33.8$, and $48.2$ are clearly visible in Fig. \ref{fig4}. The $\tilde{\gamma}=0$ energy of the lowest level of each group can be obtained only approximately from eq. (\ref{act}): due to the energy and angular momentum discretization, we have that $\lambda_1$ in $\tilde{E}= 2\left({{m_1}/ {\lambda_1}} \right)^2$ is not fixed, it varies -for scaled energies up to $\tilde{E}=140$- between $0.81$ and $1.21$. We thus have for some $m_1$ values two groups of levels, both associated with the main resonance island: for $m_1=2$ these are at $\lambda_1=1.20502$ ($\tilde{E}=5.5$) and $0.81379$ ($\tilde{E}=12.0$); for $m_1=3$, at $\lambda_1=0.91544$ ($\tilde{E}=21.5$) and $0.73098$ ($\tilde{E}=33.8$); and for $m_1=6$, at $\lambda_1=0.91653$ ($\tilde{E}=85.7$) and $0.81478$ ($\tilde{E}=108.4$); note that, again from eq. (\ref{act}), the lowest levels within each $m_1 = const$ series are those at highest $\lambda_1$.

In Fig. \ref{fig4}, the two straight lines at $\gamma=1/2\mu l=1/2$ (lower line) and at $\gamma=0.11$, where the last invariant torus disappears \cite{cla}, (upper line) mark the region of classical ``global chaos". This ``global chaos" region appears darker than the rest of the $\{\tilde{\gamma},\tilde{E}\}$ plane because it is there that the $m_1$, $-m_1$ degenerate levels significantly separate and the resulting high density of distint levels produces multilevel interactions of states corresponding to different classical resonances. These multilevel interactions correspond to the overlap of classical resonances and therefore are the quantum mechanical mark of ``global chaos" \cite{uzer}.

Two-level interactions can be classified into two types: the first one is the splitting of degenerate or near-degenerate levels increasingly repelling each other with growing $\tilde{\gamma}$, which can be locally described by the Demkov model \cite{demk}: a two level Hamiltonian with constant diagonal terms and off-diagonal terms which depend on the perturbation ($\tilde{\gamma}$ in the present case). The second type of level interaction is instead the avoided crossing, best described by the Landau-Zener model \cite{laze}: again a two level Hamiltonian, where now are the diagonal terms which depend on the perturbation while the off-diagonal terms are constant.

Demkov-like level interactions are localized either at $\tilde{\gamma}=0$ (interaction of near-degenerate resonance island levels) or in the ``global chaos" region (breaking of the $m_1$, $-m_1$ degeneracy). Landau-Zener-like interactions are instead evident almost everywhere in Fig. \ref{fig4}. Still, the highest density of both splittings and avoided crossings is in the ``global chaos" triangle, where the splitting of the $m_1$, $-m_1$ degeneracy also induces a high number of avoided crossings.

\subsubsection{Husimi Functions at Zero Gravity}

At $\tilde{\gamma}=0$, the levels are all degenerate in pairs (except the $m_1=0$ ones); the chosen basis set then decides how the probability is divided between the two states of each pair: looking at the two $\varphi_2=0$, $(\lambda_1, \varphi_1)$ planes (both $\dot{\varphi_2} >0$ and  $\dot{\varphi_2} <0$), the Husimi function of the $m_1>0$ level of a degenerate pair has support on the upper half of the plane ($\lambda_1>0$) while the Husimi of the $m_1<0$ one has support on the lower half. The size of the projection on each of the two planes instead depends on the underlying classical phase space structure: if $|m_1|<\sqrt{\tilde{E}}$ we classically have rotation in $\varphi_2$; one of the levels of the pair has therefore support either on the $\dot{\varphi_2} >0$ plane or on the $\dot{\varphi_2} <0$ one, with only a negligible tail on the other plane that becomes larger for states whose support is close to the sepatatrix; the other level vice versa. If instead $|m_1|>\sqrt{\tilde{E}}$, classically  we have libration in $\varphi_2$; both the degenerate functions therefore have significant projections on both the planes. Levels with the same $m_1$ but with support one on $\dot{\varphi_2} >0$, the other on $\dot{\varphi_2} <0$, are {\it not} degenerate.

\subsubsection{Husimi functions at Low Classical Gravity Parameter}

For $ \gamma\sim> 0$ the pairs of levels are still essentially degenerate, but the interaction -though small- begins to mix states with different $m_1$ quantum numbers; in particular there is some flow of probability between states with opposite $m_1$'s: again looking at two (quasi)-degenerate states, the tail of the $m_1>0$ Husimi now also has a component in the lower half of the SOS; likewise, the tail of the $m_1<0$ Husimi has a component in the upper half of the SOS. These tails -which for levels $410$ and $411$ are at $\tilde{\gamma} =0.025$ ($ \gamma= 2.3\cdot 10^{-4}$) already larger than the $\tilde{\gamma}= 0$ ones by five orders of magnitude- reflect the appearance of classical libration motion in $\varphi_1$ when $\gamma>0$; but, since they are due to a probability flow which happens via tunneling through the unbroken tori around $\lambda_1=0$, they remain small till the ``global chaos" triangle is reached and those tori are broken.

For $\gamma$ small enough that classical chaos is not yet global, and the resonance islands still cover most of the phase space ($\gamma<\sim0.1$), the widest avoided crossings undergone by the main resonance states are with states having similar $\lambda_1$ but with the sign changed; no probability flow is visible around $\lambda_1=0$, again because of the unbroken KAM tori in that region; the flow takes instead place between the $\dot{\varphi_2} >0$ husimi of one state and the $\dot{\varphi_2} <0$ one of the other. Three examples are given in Figs. \ref{fig5}, \ref{fig6}, and \ref{fig7}, together with the classical Poicar\'e section for the parameters of the crossing; since the functions are symetric for $\varphi_1 \rightarrow -\varphi_1$, only half of the SOS is shown; the avoided crossings of the other states of each of the degenerate doublets are identical to those shown. 

The first example (Fig. \ref{fig5}) shows a very clean avoided crossing undergone by the third state of the group of states associated with the main classical resonance originating at $\tilde{E}\simeq 86$; the parameters of the crossing place it out of Fig. \ref{fig4}, but it has been chosen because -being at higher energy- the Husimi functions are better localized in phase space and flows at the crossing can be better recognized. Both levels being in the central region of the SOS (rotation in $\varphi_2$ for $\gamma=0$) the flow is completely perpendicular to the $\{\lambda_1,\varphi_1\}$ plane: the two structures grow and fade but do not touch in the plane. 

The second example (Fig. \ref{fig6}) shows a similar avoided crossing for the first state of the same grouping; here part of the support of the second function ($b$) before the crossing is on the unstable fixed point of the principal resonance (it is therefore, at least in part a ``scarred" state \cite{hell}); to the process already seen in the previous example is thus added a visible probability flow between the main resonance island and the scarred portion of the second state. 

Finally, the third example (Fig. \ref{fig7}) shows the avoided crossing of three levels, one of them being the first state of the group of states associated with the main classical resonance originating at $\tilde{E}\simeq 108.5$. Of the other two states, one is its symmetric in $\lambda_1$ (apparently a scarred state centered on the unstable fixed point of the most noticeable resonance in the lower part of the SOS), the other one (the intervening state) is instead a mixture of a scarred state of the period two resonance above the principal one with an excited state of -again- the most noticeable resonance in the lower part of the SOS. Again, even though the support of a state can move to $\lambda_1>0$ to $\lambda_1<0$ when passing the avoided crossing, all visible flow in the $\{\lambda_1,\varphi_1\}$ plane is among structures with the same sign of $\lambda_1$. 

From our study of the Husimi functions accessible to my simulations in this regime, it appears that level interactions which can be described by the Demkov model (level splitting) take place in the $\{\lambda_1,\varphi_1\}$ plane, while interactions to be described by the Landau-Zener one (avoided crossing) thake place perpendicular to it. The first part of the above statement could be expected from what we have already seen -namely that the Demkov transitions are those mixing states with different values of $m_1$- the second part is instead a consequence of the mixing of states with different values of $n_2$ by the Landau-Zener transitions.  

\subsubsection{Husimi Functions in the ``Global Chaos" region}

As we have seen, the two processes responsible for the high density of avoided crossings in the ``global chaos" triangle are the growth in energy of the low lying states and the splitting of the $\pm m_1$ degeneracy. Both these processes are not associated with the appearance of resonance islands in the $\{\lambda_1,\varphi_1\}$ SOS as these latter are connected with the interaction of states with similar $m_1/\sqrt{\tilde{E}}$ ratios but different $|m_1|$ and $n_2$ values. 

At such $\gamma$ values the phase space is mostly taken by the chaotic sea and little remains of the classical resonant structures visible at lower $\gamma$ values but many quantum states still have Husimi functions peaked on their stable and unstable fixed points. On the other hand, due to the multiple level interactions we have already mentioned, only rarely the Husimi fuctions of states in this regime are peaked on single structures: even away from avoided crossings the support of most states covers several classical structures, resulting in rather complicated multi-peaked Husimi functions. A few examples are shown in Fig. \ref{fig8} $a$ through $d$. Fig. \ref{fig8} $e$ and $f$ instead show another typical shape for Husimi functions in this regime: the probability is concentrated along the border of the accessible classical region. Here -with the exception of the main resonance island which disappears at $\tilde{\gamma}\approx 2.0$- are the last island chains to be eaten up by the chaotic sea (at $\tilde{\gamma}\approx 1.2$) and the first ones to appear (at $\tilde{\gamma}\approx 2.5$) when with increasing $\tilde{\gamma}$ the phase space reverts to regular. 

\subsubsection{Husimi Functions at High Classical Gravity Parameter: Coupled Oscillators Regime}

For arbitrarily small quantum gravity parameter the ground state is in the ``global chaos" region and its Husimi function extends on the whole SOS; when gravity is increased it soon leaves the ``global chaos" region (at $\tilde{\gamma}=0.075$, $\gamma$ is already bigger than $1/3$) and at first concentrates on the fixed point just below the center of the $\gamma=1$ SOS in Fig. $8$ of Ref. \cite{cla}; but when -at $\tilde{\gamma}\approx 8$- it enters the coupled oscillators regime ($\gamma\sim> 2$) where the SOS consists of concentric curves, it splits in two peaks located at $\lambda_1=0$ and $\varphi_1$ close to the extreme values $\pm \arccos{[1- (3\gamma)^{-1}]}$ (inversion points of the classical orbit). 
This reflects what can be observed in Fig. $8$ of Ref. \cite{cla}: when with increasing $\gamma$ the system leaves the ``global chaos" region, the first regular structures to appear are at the rim of the SOS. 
For $\gamma\sim> 2$ both the extreme values of $\varphi_1$ and the positions in $\varphi_1$ of the peaks decrease as $1/\sqrt{\gamma}$ (or -equivalently- $\tilde{\gamma}^{-1/4}$) while the relative width of the peaks reduces.

A similar behaviour, but at much higher gravity (already the first excited state leaves the ``global chaos" region at $\tilde{\gamma}=0.625$ and enters the coupled oscillators regime at $\tilde{\gamma}=27$) and moreover complicated by avoided crossings which cause deviations from this pattern at some $\tilde{\gamma}$ values, is observed for the other states with low quantum numbers.

\section{Conclusions, Prospectives and Aknowledgements}

I have conducted the first extensive study of the dynamics of the quantum double pendulum: even if not exahustive it has allowed us to observe a close correspondence between classical and quantum structures in phase space in all three the classical regimes: from the ``free rotors" one at low classical gravity parameter $\gamma$ to the ``coupled harmonic oscillators" regime at high $\gamma$, all through the ``global chaos" regime for intermediate values of $\gamma$. In particular, notwithstanding the persistence of some regular Husimi functions in the ``global chaos" regime, the Husimi functions of most of the states in that region are quite complicated, suggesting that the time evolution of quantum packets might simulate rather well the chaotic classical evolution, spreading rapidly over most of the phase space and remaining  for fairly long times in such a state before eventually collapsing again in a localized packet, as expected from the well known analysis of Ref. \cite{ford}. The study of the time evolution of suitably placed minimum uncertainty packets will be the subject of a forthcoming paper. 

Projections of the Husimi functions on other phase space sections -$(\lambda_1, \varphi_1)$ planes for $\varphi_2\neq 0$ and $(\lambda_2, \varphi_2)$ planes for different values of $\varphi_1$- and investigation of other combinations of the length and mass parameters $l$ and $\mu$ might give some interesting insight too, especially when compared to the case studied here. 

I wish to thank G. Mantica and S. Locklin for useful comments and suggestions.

\newpage

\begin{figure}[htbp]
\centering\epsfig{file=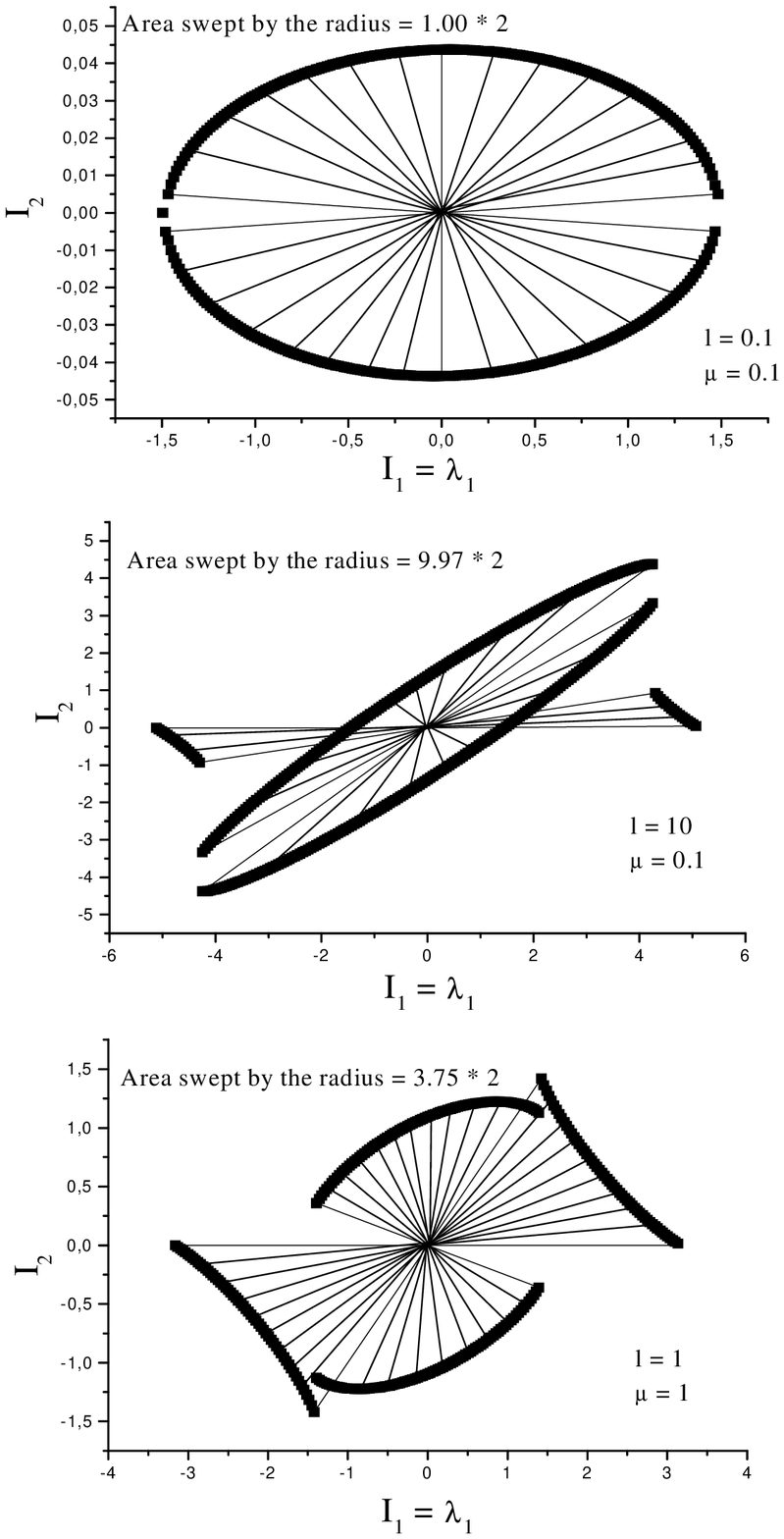,width=1.0\linewidth}
\caption{Energy surfaces in scaled action variable representation.The plots show $I_2$ versus $I_1$ at constant energy $h=1$. The radii indicate the areas to be calculated to evaluate the density of states.}
\label{fig2}
\end{figure}

\begin{figure}[htbp]
\centering\epsfig{file=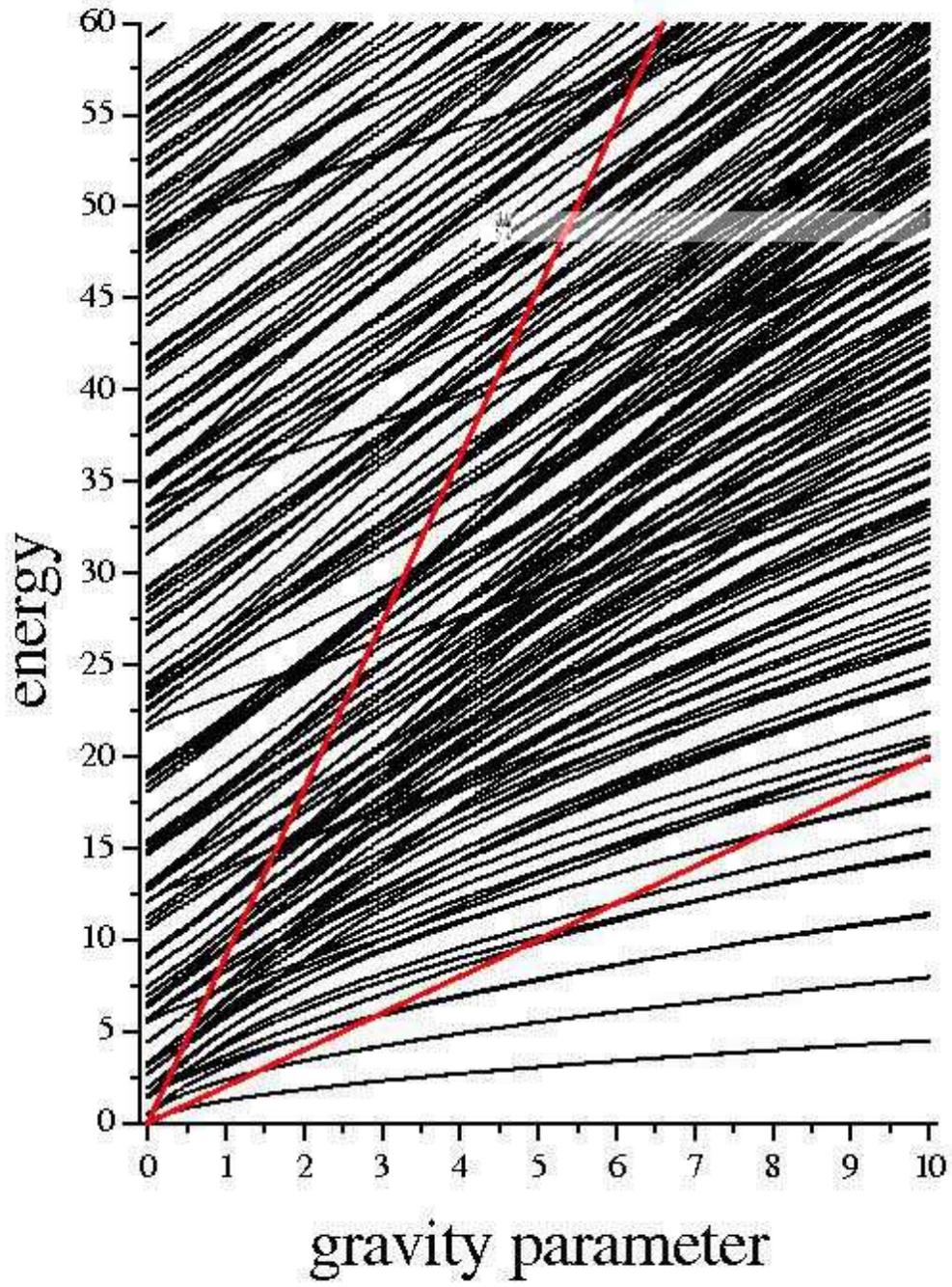,width=0.9\linewidth}
\caption{Energy curves.}
\label{fig4}
\end{figure}

\begin{figure}[htbp]
\centering\epsfig{file=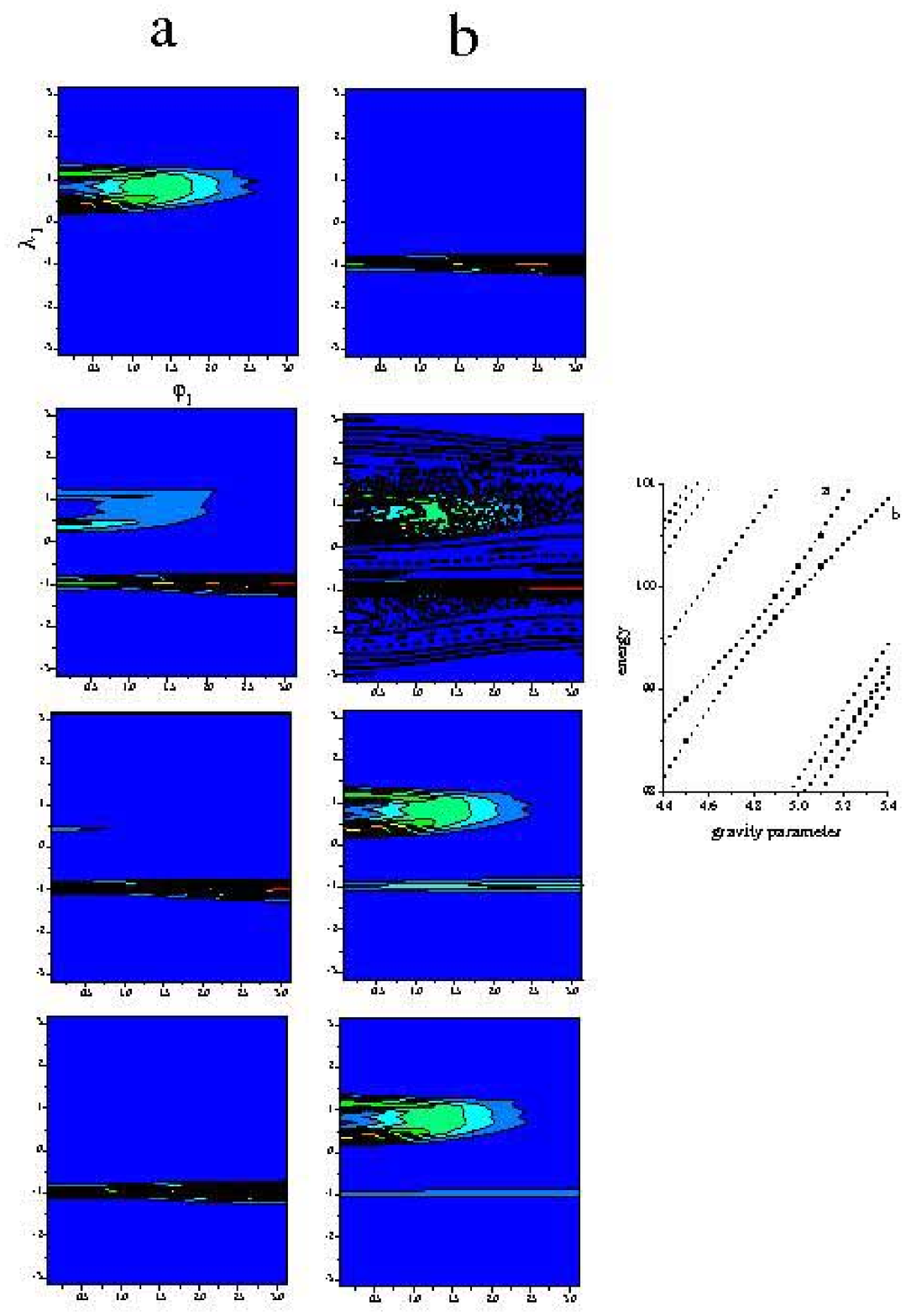,width=0.9\linewidth}
\caption{(Color online) An avoided crossing of two levels, one of which is a principal resonance one. The points at which the Husimi functions are calculated are marked as bigger dots on the energy curves shown on the right of the figure. The gravity parameter increases from top to bottom. For comparison, the classical Poincar\'e section for $\gamma=0.04888$ is superimposed on the corresponding Husimi function.}
\label{fig5}
\end{figure}

\begin{figure}[htbp]
\centering\epsfig{file=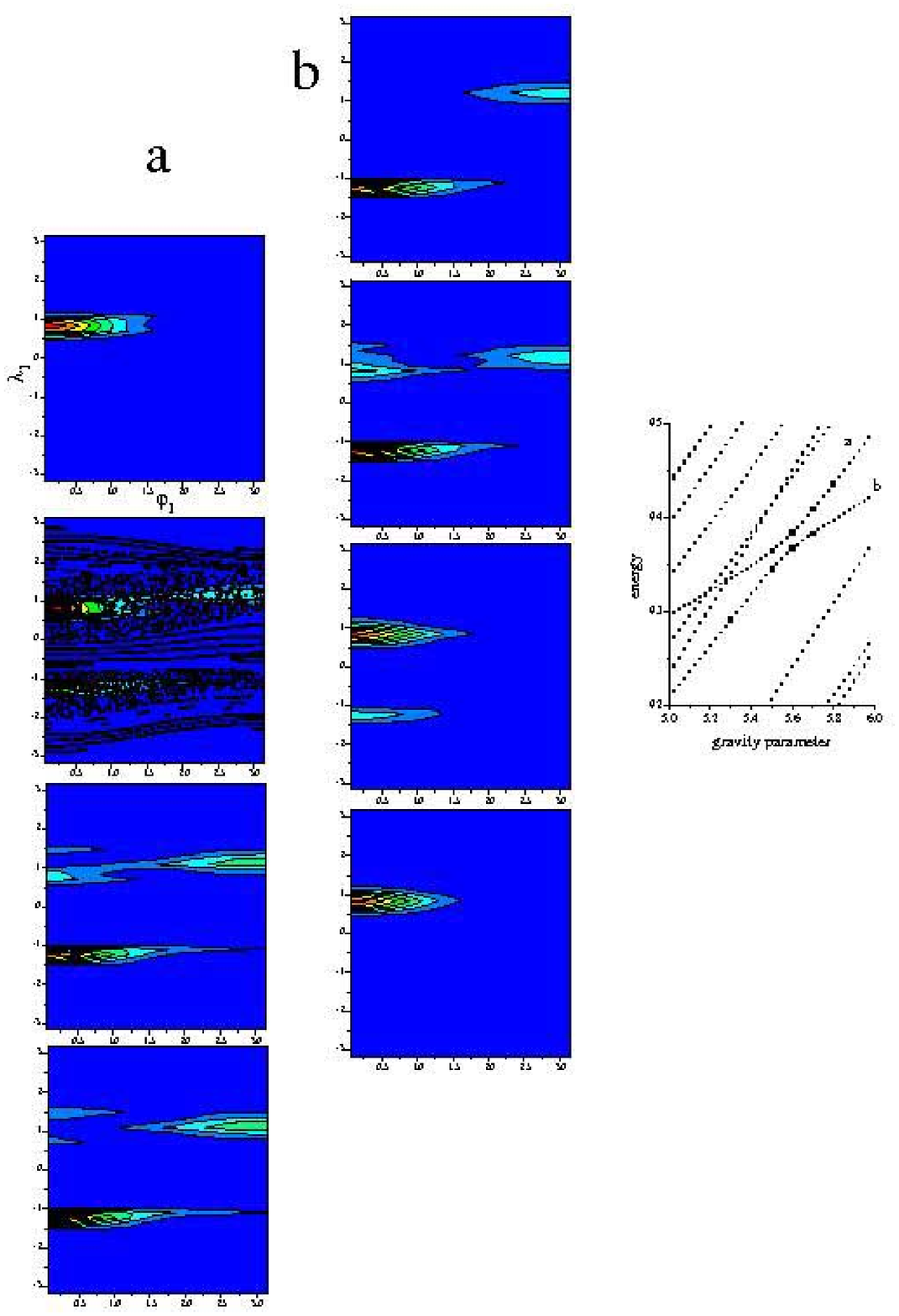,width=0.9\linewidth}
\caption{(Color online) An avoided crossing of three levels, one of which is a principal resonance one. The points at which the Husimi functions are calculated are marked as bigger dots on the energy curves shown on the right of the figure. The gravity parameter increases from top to bottom. For comparison, the classical Poincar\'e section for $\gamma=0.0597877$ is superimposed on the corresponding Husimi function.}
\label{fig6}
\end{figure}

\begin{figure}[htbp]
\centering\epsfig{file=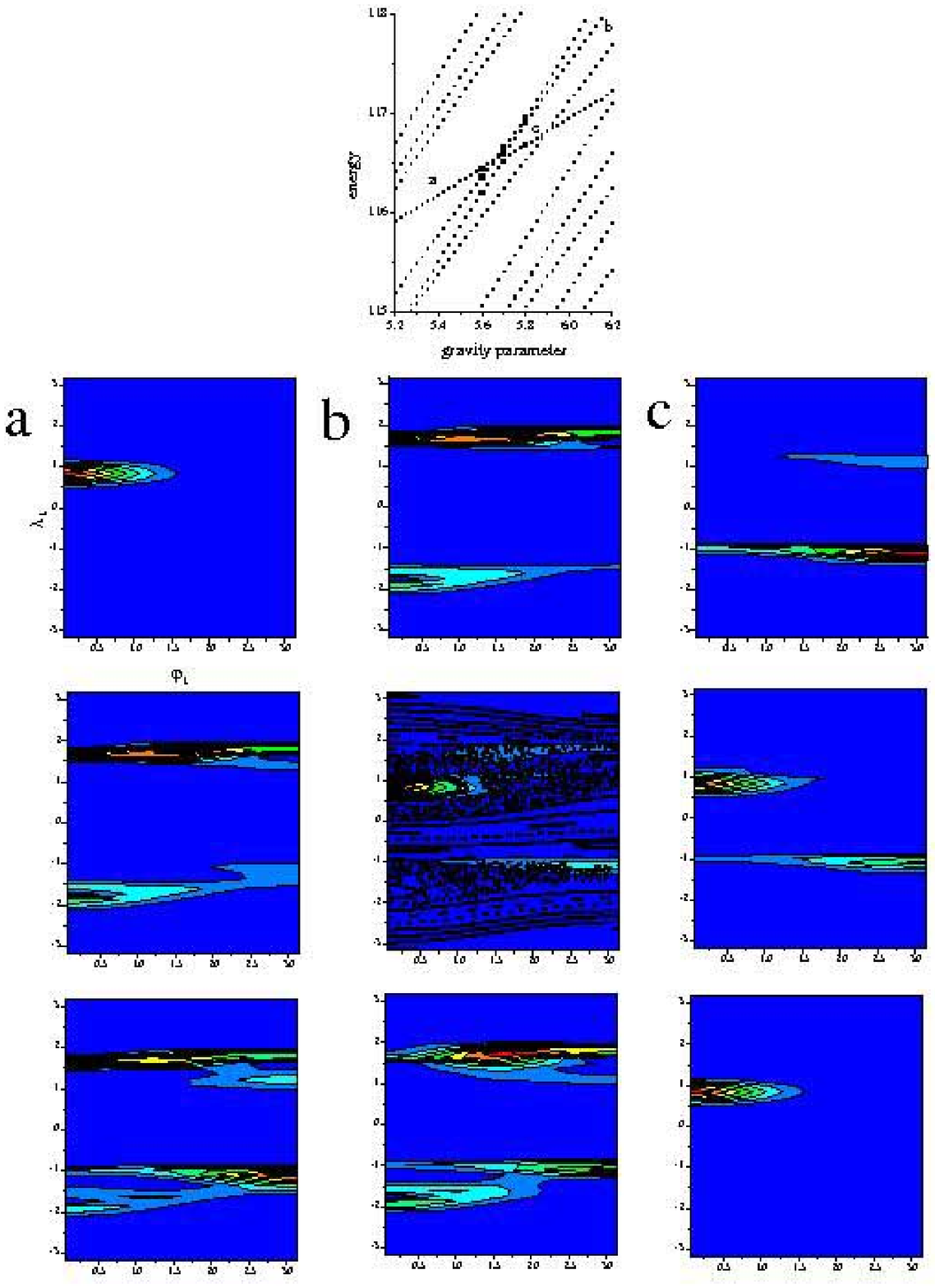,width=0.9\linewidth}
\caption{(Color online) An avoided crossing of two levels, one of which is a principal resonance one. The points at which the Husimi functions are calculated are marked as bigger dots on the energy curves shown on top of the figure. The gravity parameter increases from top to bottom. For comparison, the classical Poincar\'e section for $\gamma=0.04888$ is superimposed on the corresponding Husimi function.}
\label{fig7}
\end{figure}

\begin{figure}[htbp]
\centering\epsfig{file=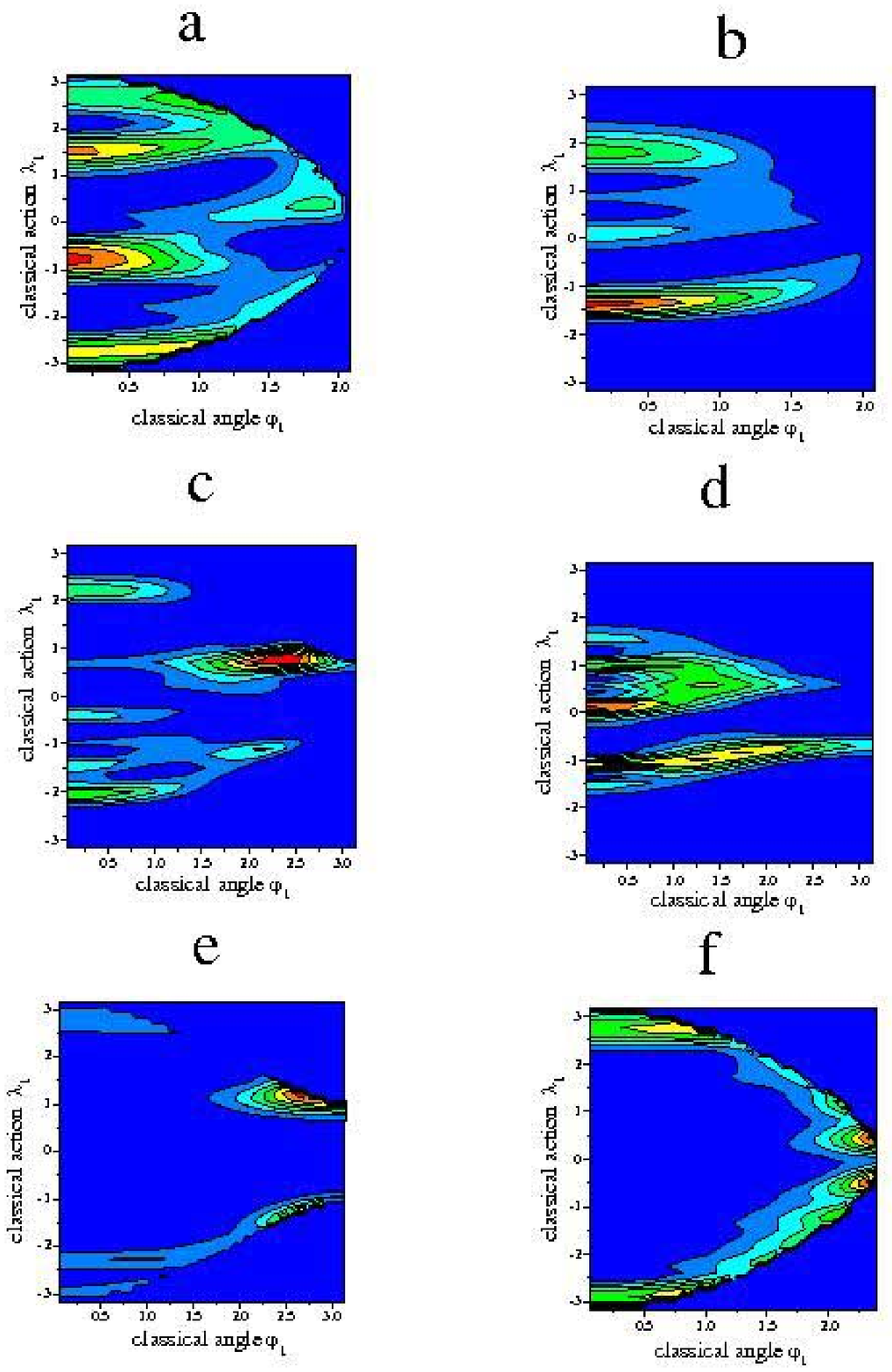,width=0.9\linewidth}
\caption{(Color online) Examples of Husimi functions in the ``global chaos" region. {\it a)}: $n=33$, $\tilde{\gamma}=5.4$, $\gamma=0.22445$. {\it b)}: $n=36$, $\tilde{\gamma}=5.4$, $\gamma=0.21950$. {\it c)}: $n=127$, $\tilde{\gamma}=9.8$, $\gamma=0.15692$. {\it d)}: $n=118$, $\tilde{\gamma}=8.1$, $\gamma=0.14632$. {\it e)}: $n=117$, $\tilde{\gamma}=8.1$, $\gamma=0.14815$. {\it f)}: $n=74$, $\tilde{\gamma}=8.6$, $\gamma=0.19053$.}
\label{fig8}
\end{figure}


\begin{references}
\bibitem{note4} For convenience in our numerical simulations, we introduce the composite index $j=(m_1 + [N_1/2])N_2+(m_2+ [N_2/2])$, where $N_1 , N_2$ are the numbers of rotor states taken for each of the two spatial variables (odd for symmetry reasons), and $[\hspace{.1in} ]$ denotes integer part. The kinetic energy matrix is then block diagonal with $N_1$ square blocks of dimension $N_2\times N_2$. The potential energy is instead band diagonal with only the four diagonals $\{j,j\pm N_2\}$ and $\{j,j\pm (N_2+1)\}$ different from zero.
\bibitem{note5} We note here that the symmetry of the constant energy curves in the $(I_1,I_2)$ plane (see Fig. \ref{fig2}) means a double degeneracy of all levels that is compatible with the uniqueness of the ground state only if the Maslov indices inside the separatrix (rotation in both angles) are zero, in agreement with the chosen symmetrization in eq. (\ref{sch}).
\end{references}
\end{document}